\documentclass[10pt]{article}
\usepackage{amsmath}
\usepackage{amssymb}
\usepackage{graphicx}

\def\aa{A\&A }

\def\aj{AJ }

\def\mnras{MNRAS }

\begin{document}

\setcounter{figure}{0}
\setcounter{table}{0}
\setcounter{footnote}{0}
\setcounter{equation}{0}

\vspace*{0.5cm}
\noindent {\Large A VLBI SURVEY OF WEAK EXTRAGALACTIC RADIO SOURCES FOR THE ALIGNMENT OF THE ICRF AND THE FUTURE GAIA FRAME}
\vspace*{0.7cm}

\noindent\hspace*{1.5cm} G. BOURDA$^1$, P. CHARLOT$^1$, R. PORCAS$^2$, S. GARRINGTON$^3$ 
\vspace{0.2cm}

\noindent\hspace*{1.5cm} $^1$ Laboratoire d'Astrophysique de Bordeaux (LAB),       \\
\noindent\hspace*{1.5cm} ~~Observatoire Aquitain des Sciences de l'Univers (OASU), \\ 
\noindent\hspace*{1.5cm} ~~Universit\'{e} Bordeaux 1, CNRS                         \\
\noindent\hspace*{1.5cm} ~~BP89, 33270 Floirac, France                             \\
\noindent\hspace*{1.5cm} ~~e-mail: Geraldine.Bourda@obs.u-bordeaux1.fr                       
\vspace{0.15cm}

\noindent\hspace*{1.5cm} $^2$ Max Planck Institute for Radio Astronomy  \\
\noindent\hspace*{1.5cm} ~~P.O. Box 20 24, 53010 Bonn, Germany                               
\vspace{0.15cm}

\noindent\hspace*{1.5cm} $^3$ Jodrell Bank Observatory, The University of Manchester  \\
\noindent\hspace*{1.5cm} ~~Macclesfield, Cheshire, SK11 9DL, UK.                      \\

\vspace*{0.5cm}
\noindent {\large ABSTRACT.} The space astrometry mission GAIA will construct a dense optical QSO-based celestial reference frame. For consistency between the optical and radio positions, it will be important to align the GAIA frame and the International Celestial Reference Frame (ICRF) with the highest accuracy. Currently, it is found that only 10\% of the ICRF sources are suitable to establish this link, either because they are not bright enough at optical wavelengths or because they have significant extended radio emission which precludes reaching the highest astrometric accuracy. In order to improve the situation, we have initiated a VLBI survey dedicated to finding additional high-quality radio sources for aligning the two frames. The sample consists of about 450 sources, typically 20 times weaker than the current ICRF sources, which have been selected by cross-correlating optical and radio catalogues. The paper presents the observing strategy and includes preliminary results of observation of 224 of these sources with the European VLBI Network in June 2007.

\vspace*{1cm}
\noindent {\large 1. CONTEXT}
\smallskip

The ICRF (International Celestial Reference Frame; Ma~et~al.~1998; Fey~et~al.~2004) is the fundamental celestial reference frame adopted by the International Astronomical Union (IAU) in August 1997. It is currently based on the VLBI (Very Long Baseline Interferometry) positions of 717 extragalactic radio sources, estimated from dual-frequency S/X (2.3 and 8.6 GHz) observations. 
The European space astrometry mission GAIA, to be launched by 2011, will survey about one billion stars in our Galaxy and 500~000 Quasi Stellar Objects (QSOs), brighter than magnitude 20 (Perryman~et~al.~2001). Unlike Hipparcos, GAIA will construct a dense optical celestial reference frame directly in the visible, based on the QSOs with the most accurate positions (i.e. those with optical apparent magnitude $V\leq18$; Mignard~2003). 
In the future, the alignment of the ICRF and the GAIA frame will be crucial for ensuring consistency between measured radio and optical positions. This alignment, to be determined with the highest accuracy, requires hundreds of sources in common, with a uniform sky coverage and very accurate radio and optical positions. Obtaining such accurate positions implies that the link sources must have $V\leq18$ and no extended VLBI structures. 

In a previous study, we investigated the current status of this link based on the present list of ICRF sources (Bourda~et~al.~2008). We found that although about 30\% of the ICRF sources have an optical counterpart with $V \leq 18$, only one third of these are compact enough on VLBI scales for the highest astrometric accuracy. 
Overall, only 10\% of the current ICRF sources ($\simeq$70~sources) are thus available for the alignment with the GAIA frame. This highlights the need to identify additional suitable radio sources, which is the purpose of the project described here.

\newpage
\noindent {\large 2. STRATEGY TO IDENTIFY NEW LINK RADIO SOURCES}
\smallskip

Searching for additional radio sources suitable for aligning accurately the ICRF and the GAIA frame implies going to weaker radio sources having flux densities typically below 100~mJy. This can now be realized owing to recent increases in the VLBI network sensitivity (e.g. recording now possible at 1Gb/s) and by using a network comprising large antennas like the European VLBI Network (EVN).

A sample of about 450 radio sources, for which there are no published VLBI observations, was selected for this purpose by cross-identifying the NRAO VLA Sky Survey (NVSS; Condon~et~al.~1998), a deep radio survey (complete to the 2.5~mJy level) which covers the entire sky north of $-40^{\circ}$, with the V\'{e}ron~\&~V\'{e}ron~(2006) optical catalogue. This sample is based on the following criteria: $V\leq18$ (to ensure very accurate positions with GAIA), $\delta\geq-10^{\circ}$ (for possible observing with northern VLBI arrays), and NVSS flux density~$\geq$~20~mJy (for possible VLBI detection). 

The observing strategy to identify the most appropriate link sources from this sample includes three successive steps to detect, image and measure accurately the astrometric position of these sources: (i) to determine the VLBI detectability of these weak radio sources; (ii) to image and determine an accurate astrometric position for the sources detected in the first step; and (iii) to refine the astrometry for the most compact sources of the sample.

\vspace*{0.7cm}
\noindent {\large 3. INITIAL VLBI RESULTS}
\smallskip

The first observations for this project (experiment EC025A) were carried out in June 2007, with a network of four EVN telescopes (i.e. Effelsberg, Medicina, Noto, Onsala). Half of our sample (i.e. 224~target sources, most of which belong to the CLASS catalogue; Myers~et~al.~2003) was observed during this experiment to determine their VLBI detectability (step 1 described above). The rest of the sources were observed in October 2007. 

\begin{figure}[h]
\begin{center}
  \includegraphics[scale=0.31,angle=-90]{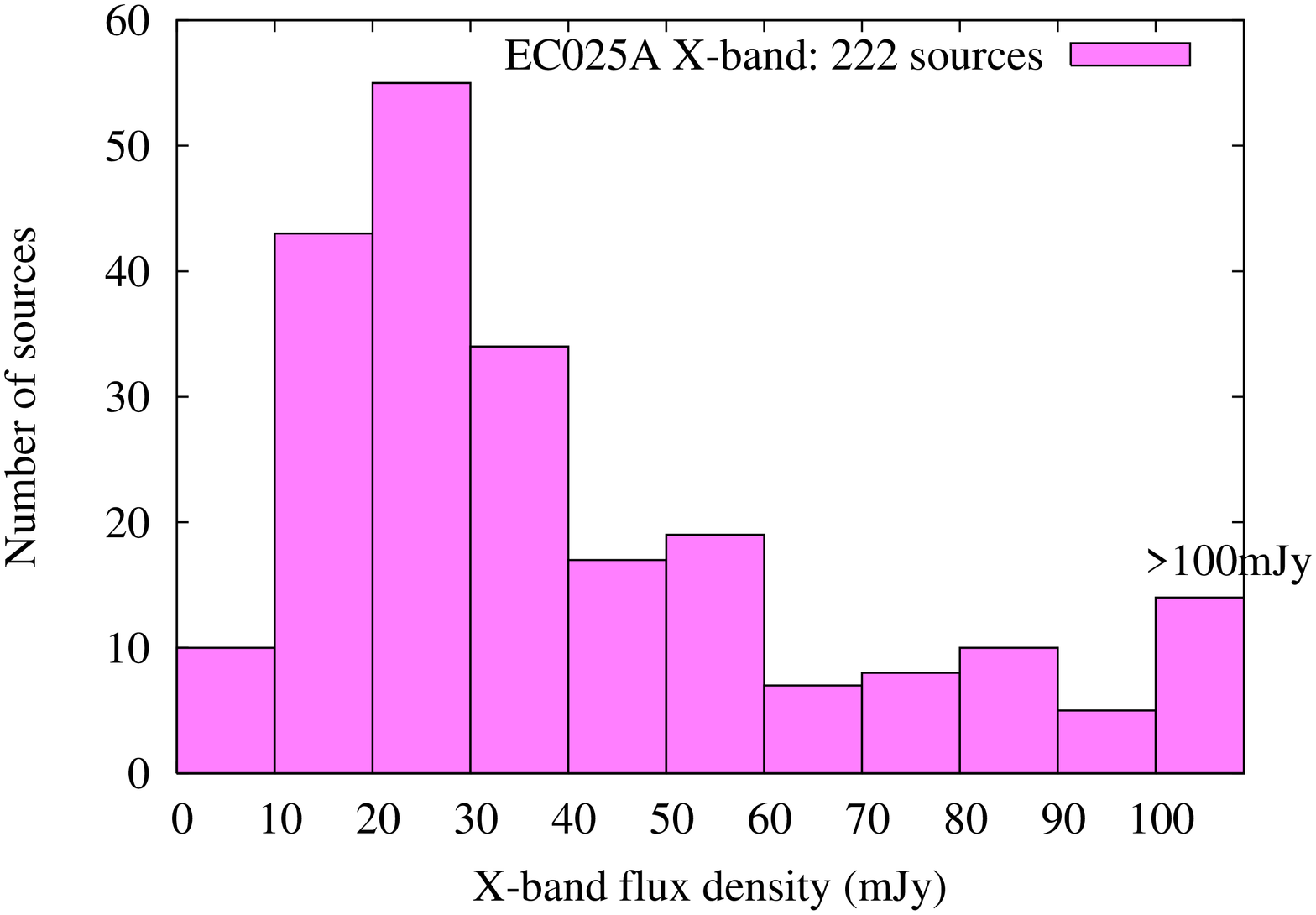}
  \includegraphics[scale=0.31,angle=-90]{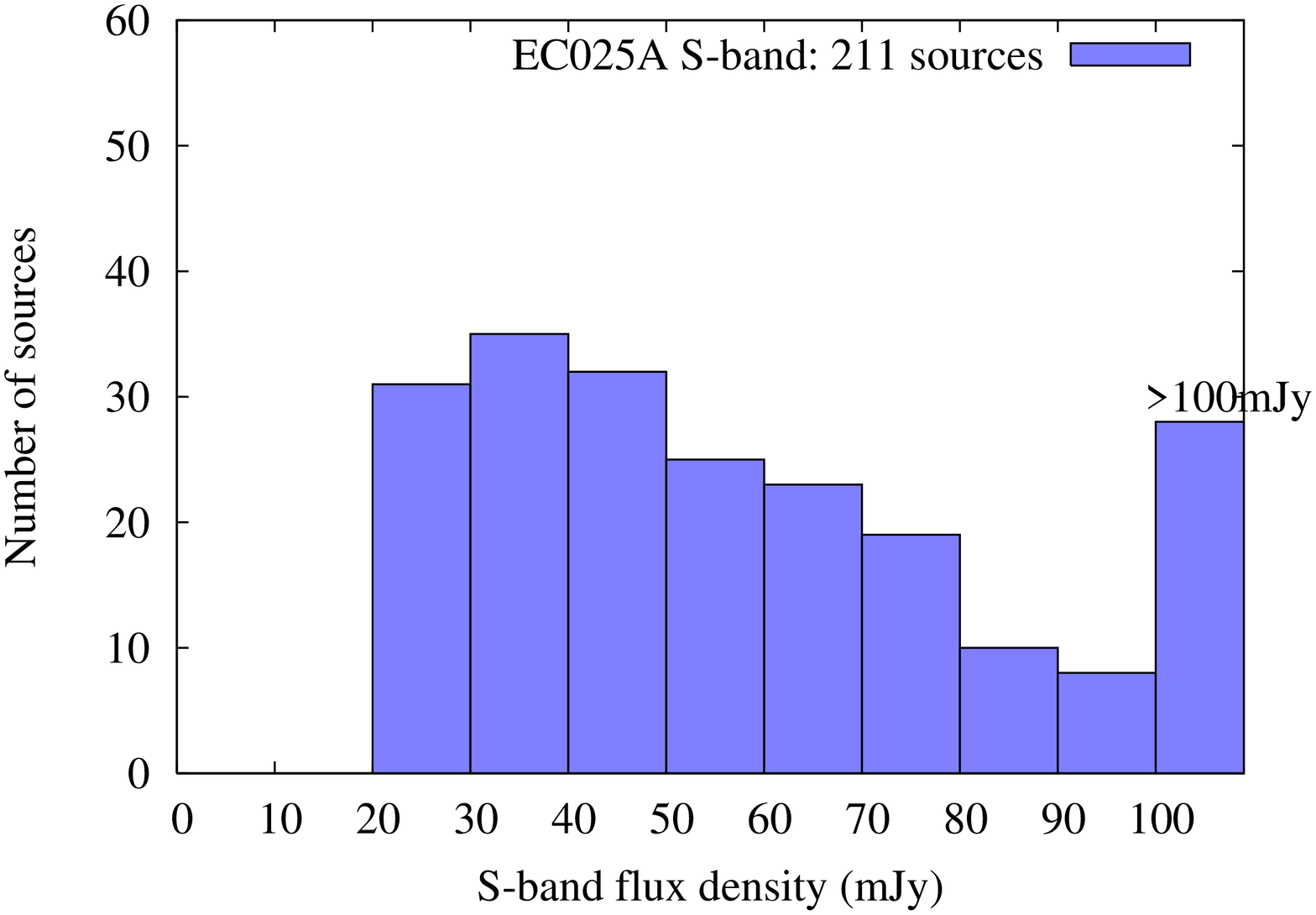}
\caption{Mean correlated flux density distribution (units in mJy), at X band and S band, for the sources detected in our initial experiment (EC025A) conducted in June 2007.}\label{fig:Fig1}
\end{center}
\end{figure}

\begin{figure}[hp]
\begin{center}
  \includegraphics[scale=0.7]{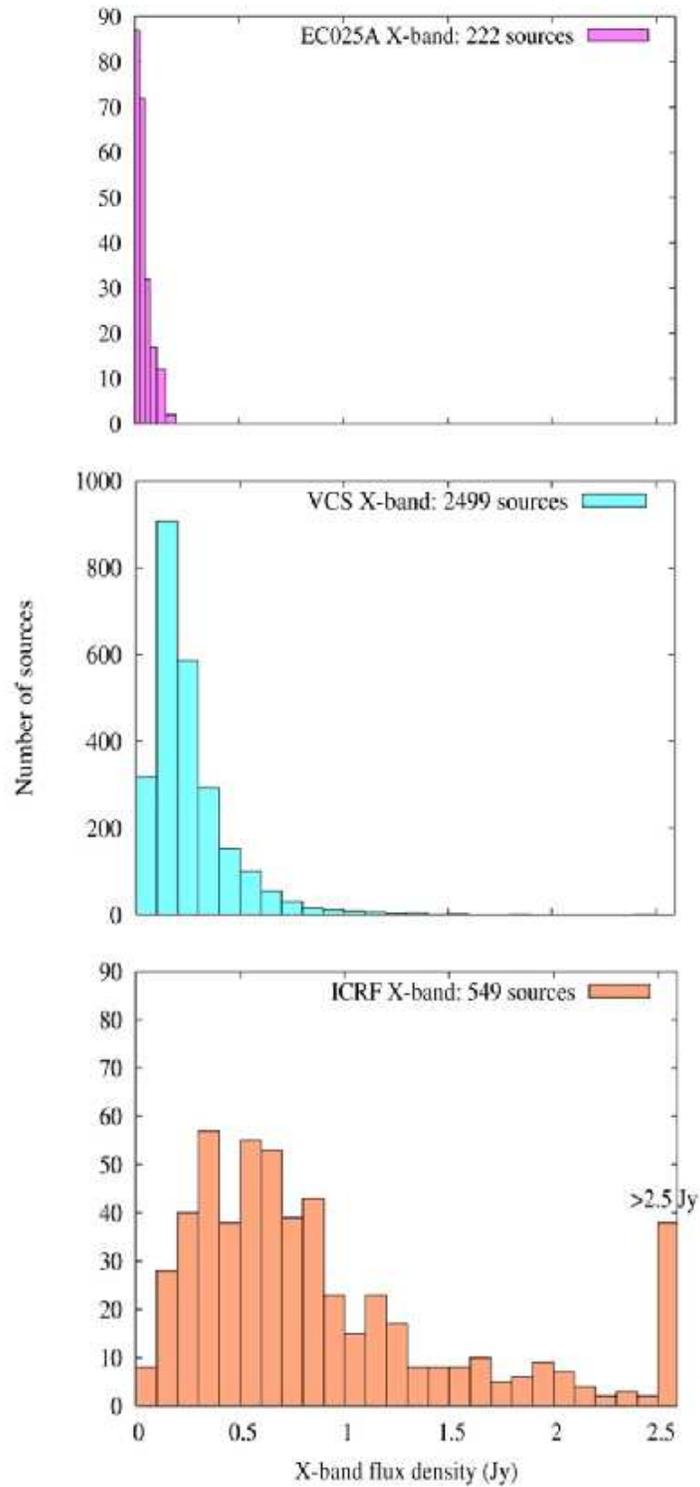}
\caption{Comparison of the X-band flux density distribution (units in Jy) for the sources detected in EC025A and those from the VCS and ICRF catalogues.}\label{fig:Fig2}
\end{center}
\end{figure}

Our results for EC025A indicate excellent detection rates of 99\% at X band and 95\% at S band, with 222~sources and 211~sources detected at X and S bands, respectively. The mean correlated flux densities have a median value of 32~mJy at X band and 55~mJy at S band (see Figure~\ref{fig:Fig1}). In Figure~\ref{fig:Fig2}, the comparison of the X-band flux density distribution for the EC025A, VCS (Beasley et al. 2002; Fomalont et al. 2003; Petrov et al. 2005, 2006; Kovalev et al. 2007) and ICRF sources shows that the sources of our sample are indeed much weaker. The average flux density of the sources detected in EC025A is 20~times and 7~times weaker than those of the ICRF and VCS sources, respectively.  

The spectral index $\alpha$ of the 211 radio sources detected at both frequencies was also investigated. This parameter is defined by $S \propto \nu^{\alpha}$, where $S$ is the source flux density and $\nu$ is the frequency. Its median value is $-0.3$ and most of the sources have $\alpha > -0.5$, hence indicating that they must have a dominating core component, which is very promising for the future stages of this project.

\begin{figure}[h]
\begin{center}
  \includegraphics[scale=0.3,angle=-90]{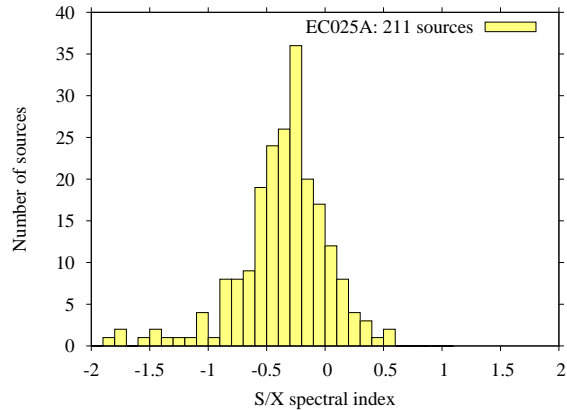}
\caption{S/X spectral index distribution for the 211 weak extragalactic radio sources detected at both S and X bands during EC025A.}\label{fig:Fig3}
\end{center}
\end{figure}

\newpage
\noindent {\large Acknowledgements}
\smallskip

\noindent This work has benefited from research funding from the European Community's sixth Framework Programme under RadioNet R113CT 2003 5058187. The authors acknowledge the EVN, which is a joint facility of European, Chinese, South African and other radio astronomy institutes funded by their national research councils. They also thank Dave Graham for assistance with the correlation in Bonn, and John Gipson for advice when scheduling the observations. The first author would like to thank the CNES (Centre National d'Etudes Spatiales, France) for the post-doctoral position granted at LAB.

\vspace*{0.7cm}
\noindent {\large 4. REFERENCES}
{

\leftskip=5mm
\parindent=-5mm

\smallskip

Bourda, G., Charlot, P., \& Le~Campion, J.-F., 2008, ``Astrometric suitability of optically-bright ICRF sources for the alignment of the ICRF with the future GAIA celestial reference frame'', \aa, submitted.

Beasley, A.J., Gordon, D., Peck, A.B., Petrov, L., MacMillan, D.S., et al., 2002, ``The VLBA Calibrator Survey-VCS1'', Astrophy. J. Supp. Series 141, pp. 13--21.

Condon, J.J., Cotton, W.D., Greisen, E.W., Yin, Q.F., Perley, R., et al., 1998, ``The NRAO VLA Sky Survey'', \aj 115, pp. 1693--1716.

Fey, A.L., Ma, C., Arias, E.F., Charlot, P., Feissel-Vernier, M., et al., 2004, ``The Second Extension of the International Celestial Reference Frame: ICRF-EXT.1'', \aj 127, pp. 3587--3608.

Fomalont, E.B., Petrov, L., MacMillan, D.S., Gordon, D., \& Ma, C., 2003, ``The Second VLBA Calibrator Survey: VCS2'', \aj 126, pp. 2562--2566.

Kovalev, Y., Petrov, L., Fomalont, E., \& Gordon, D., 2007, ``The Fifth VLBA Calibrator Survey: VCS5'', \aj 133, pp. 1236--1242.

Ma, C., Arias, E.F., Eubanks, T.M., Fey, A.L., Gontier, A.-M., et al., 1998, ``The International Celestial Reference Frame as Realized by Very Long Baseline Interferometry'', \aj 116, pp. 516--546.

Mignard, F., 2003, ``Realization of the inertial frame with GAIA'', in: R. Gaume, D. McCarthy \& J. Souchay (eds.), IAU~25~Joint~Discussion~16: The International Celestial Reference System, Maintenance and Future Realizations, pp. 133--140.

Myers, S.T., Jackson, N.J., Browne, I.W.A., de Bruyn, A.G., Pearson, T.J., et al., 2003,  ``The Cosmic Lens All-Sky Survey - I. Source selection and observations'', \mnras 341, pp. 1--12.

Perryman, M.A.C., de Boer, K.S., Gilmore, G., Hog, E., Lattanzi, M.G., et al., 2001, ``GAIA: Composition, formation and evolution of the Galaxy'', \aa 369, pp. 339--363.

Petrov, L., Kovalev, Y., Fomalont, E., \& Gordon, D., ``The Third VLBA Calibrator Survey: VCS3'', 2005, \aj 129, pp. 1163--1170.

Petrov, L., Kovalev, Y., Fomalont, E., \& Gordon, D., ``The Fourth VLBA Calibrator Survey: VCS4'', 2006, \aj 131, pp. 1872--1879. 

V\'{e}ron-Cetty, M.-P., \& V\'{e}ron, P., 2006, ``A catalogue of quasars and active nuclei: 12th edition'', \aa 455, pp. 773--777.

}

\end{document}